\pdfoutput=1
\documentclass[preprint,12pt,authoryear]{elsarticle}

\usepackage{amssymb,amsmath}

\journal{Physics Letters A}
\usepackage{graphicx}

\renewcommand{\vec}[1]{{\mathbf #1}}

\newcommand{\rmd}{ {\mathrm d} }

\newcommand{\pder}[2]{ \frac{\partial #1}{\partial #2} }

\begin{document}

\begin{frontmatter}



\title{Effects of variable thermal diffusivity on the structure of convection}


\author[ipm]{O.V. Shcheritsa}
\author[niiyaf]{A.V. Getling\corref{avg}}
\ead{A.Getling@mail.ru}
\author[ipm]{O.S. Mazhorova}
\address[ipm]{Keldysh Institute of Applied Mathematics, Moscow, 125047 Russia}
\address[niiyaf]{Skobeltsyn Institute of Nuclear Physics, Lomonosov Moscow State University, Moscow, 119991 Russia}
\cortext[avg]{Corresponding author}

\begin{abstract}
The structure of multiscale convection in a thermally stratified plane horizontal fluid layer is investigated by means of numerical simulations. The thermal diffusivity is assumed to produce a thin boundary sublayer convectively much more unstable than the bulk of the layer. The simulated flow is a superposition of cellular structures with three different characteristic scales. In contrast to the largest convection cells, the smaller ones are localised in the upper portion of the layer. The smallest cells are advected by the larger-scale convective flows. The simulated flow pattern qualitatively resembles that observed on the Sun.
\end{abstract}


\end{frontmatter}



\section{Introduction}
\label{intr}
This study was motivated by the necessity of comprehending the physical factors responsible for the complex spatial structure of solar convection. The magnetic fields in the solar convection zone are dynamically coupled with motions, and
the formation of convection patterns is of paramount importance to the dynamics of magnetic fields and, therefore, to the processes of solar activity. As is
well known, cellular flow structures of at least three or four types can be
identified with certainty on the solar surface and attributed to the phenomenon of thermal convection, viz., granules, mesogranules (whose existence as a
physical entity is debatable), supergranules and giant cells \citep[see, in
particular,][and references therein]{Nrdlnd_Stein_Asplnd_lrsp:2009,RieutRinc2010}. These different sorts of structures differ widely in their scale. Furthermore, \citet{Abramenko_etal2012} reported the detection of
mini-granular structures with spatial scales below 600~km. {The convective flow in the photospheric and subphotospheric layers is thus a superposition of these differently scaled cellular flows, so that} smaller cells are transferred by the motions associated with the larger scales. This multiscale, {hierarchical}  structure of the flow (or scale splitting) is an important
feature of solar convection. It has not yet received a convincing explanation,
and an adequate hydrodynamic description must be given to both the spatial
structure of the flows and the factors responsible for its development.

The multiscale structure of solar convection can be revealed using various observational techniques. However, the power spectra of
the velocity field seem to definitely exhibit only two peaks, which correspond to granulation and supergranulation \citep[see, e.g.,][]{Hathaway_etal2000}. The broadband character of the spectrum smears out the other peaks. Nevertheless, the structures {forming no pronounced spectral peaks} can be identified using other methods, such as local correlation tracking (LCT), wavelet analyses, tracking the motion of supergranules and an LCT-based cork-motion-tracking technique very clearly visualises both supergranules and mesogranules. A further discussion of the pros and cons concerning the existence of mesogranules is given by \citet{RieutRinc2010}.

{It is known that convection cells in various flows have typically comparable horizontal and vertical sizes. This suggests that convection structures of different types in the solar convection zone should occupy layers of different thicknesses. Since all these structures can be detected at the solar surface, it can readily be understood that they all (except the giant cells filling the whole convection-zone depth) are ``suspended'' near the upper boundary of the convection zone, while the lower boundaries of layers of different types are located at different depths.}

As for hydrodynamic modelling, some ``realistic'' \citep[or, as termed by][
``comprehensive'']{Schuessler2013} simulations of solar convection (aimed at reproducing the physical processes involved in the solar convective phenomena as closely as possible), which are based on the \emph{MURaM} code
\citep{Voegler_etal2005}, demonstrate a gradual increase in the characteristic scale of the flow with depth \citep{Schuessler2014}. According to such simulations, the surface velocity field does not demonstrate any multiscale structure, and only granular-sized cells are clearly noticeable in the computed patterns. Larger-scale structures cannot be detected at the surface and at small depths; at most, they may be too weak to manifest themselves in advecting small-scale cells. This suggests that some physics responsible, e.g., for supergranulation is not taken into account in the known versions of the ``realistic'' problems.

It is natural to believe that the structure of convection should generally be controlled by the particularities of the fluid-layer stratification. Our aim is to investigate the role of certain physical mechanisms that can give rise to a multiscale structure of convection -- first of all, by producing static temperature profiles of particular shapes in the fluid layer. At this stage of research, we do not pursue the aim of closely reproducing the physical conditions in the solar convection zone and the convection patterns actually present on the Sun and seek for physical factors capable of producing the scale-splitting effect.

Specifically, we consider here the effects of temperature-dependent thermal diffusivity. We assume this quantity to vary in such a way that the static temperature gradient $\rmd T_\mathrm s/\rmd z$ is small in the bulk of the layer but jumpwise changes to high absolute values in a thin sublayer near the top boundary of the layer. {It is important that this gradient is negative at any height; therefore, the thermal stratification is everywhere convectively unstable. This means that the phenomenon of penetrative convection has nothing to do with the subject of our study.} Such {a thermal-gradient jump resembles (although does not reproduce)} a partial-ionisation layer in the solar convection zone, where the enhanced specific heat reduces the adiabatic thermal gradient and the enhanced opacity increases the radiative thermal gradient. The convective instability of such a layer is therefore especially high \citep{SimLeighton1964,November_etal1981}. The most pronounced jump of the vertical entropy gradient due to partial ionisation is located at depths of order 1~Mm below the solar photosphere. Here, we do not claim to propose a model of the solar convection zone but merely consider the physical effects of a certain stratification peculiarity under idealised conditions, for an incompressible fluid. {Not only may this problem be of help in seeking structure-forming factors for solar convection but it is also interesting from a purely hydrodynamic standpoint.}

Linear problems of convective stability in layers with similar static
temperature profiles were considered by \citet{Getling:1975,Getling1980}. In
the framework of the incompressible-fluid models, indirect evidence for possible scale splitting was detected. The development of small-scale convective motions near the surface of the layer was found to require very sharp gradient jumps and very thin high-gradient sublayers. These expectations were partially substantiated by two-dimensional nonlinear numerical simulations of convection \citep{GetlingTikh2007}. Recently, we studied the two-dimensional problem more extensively \citep{Getling_etal2013,Shcheritsa_etal2015}.

Here, we present the results of our simulations of three-dimensional convection under conditions similar to those assumed in \citet{Shcheritsa_etal2015} in terms of the special form of the temperature dependence of thermal diffusivity. We shall demonstrate that, even in the framework of a model based on an extended Boussinesq approximation, variable thermal diffusivity can produce a multiscale flow in which at least three cell types are present.

{Our present study is hydrodynamic rather than astrophysical. Further steps toward adequately describing the flow structure in the solar convection zone should, in our opinion, include the consi\-der\-ation of different structure-forming processes in parallel with successively taking into account more physics involved.}

\section{Formulation of the problem and numerical technique}

We consider a rectangular box $[0,\, L_x]\times[0,\, L_y]\times [0,\,h]$ of
a plane horizontal layer of a viscous, incompressible fluid (in our
computations described here, $L_x=L_y=15h$). Let the bottom and the top
boundary of the layer to be perfect thermal conductors, whose temperatures are
maintained constant and equal to $T_\mathrm{bot}\equiv\Delta T>0$ and
$T_\mathrm{top}=0$, respectively. Also let the sidewalls of the region be thermally insulated. We specify the no-slip impermeability conditions at the
bottom and side boundaries of the region. The top boundary is also assumed to be rigid.

\begin{figure}
\begin{center}
{\includegraphics[width=0.56\textwidth,bb=0 0 485 330pt,clip]
{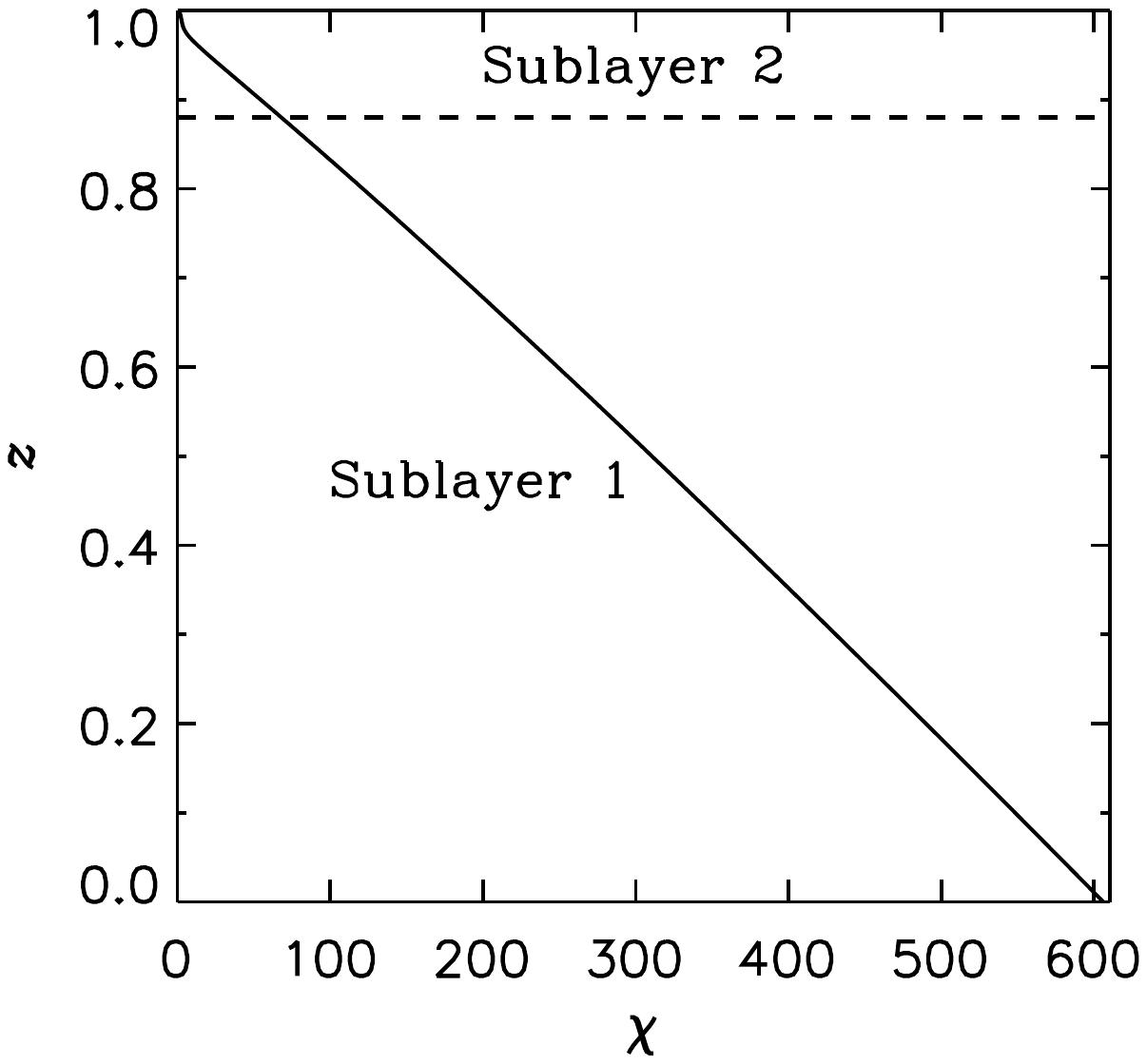}\hspace{-2cm}
\includegraphics[width=0.405\textwidth,bb=40 10 384 335pt,clip]
{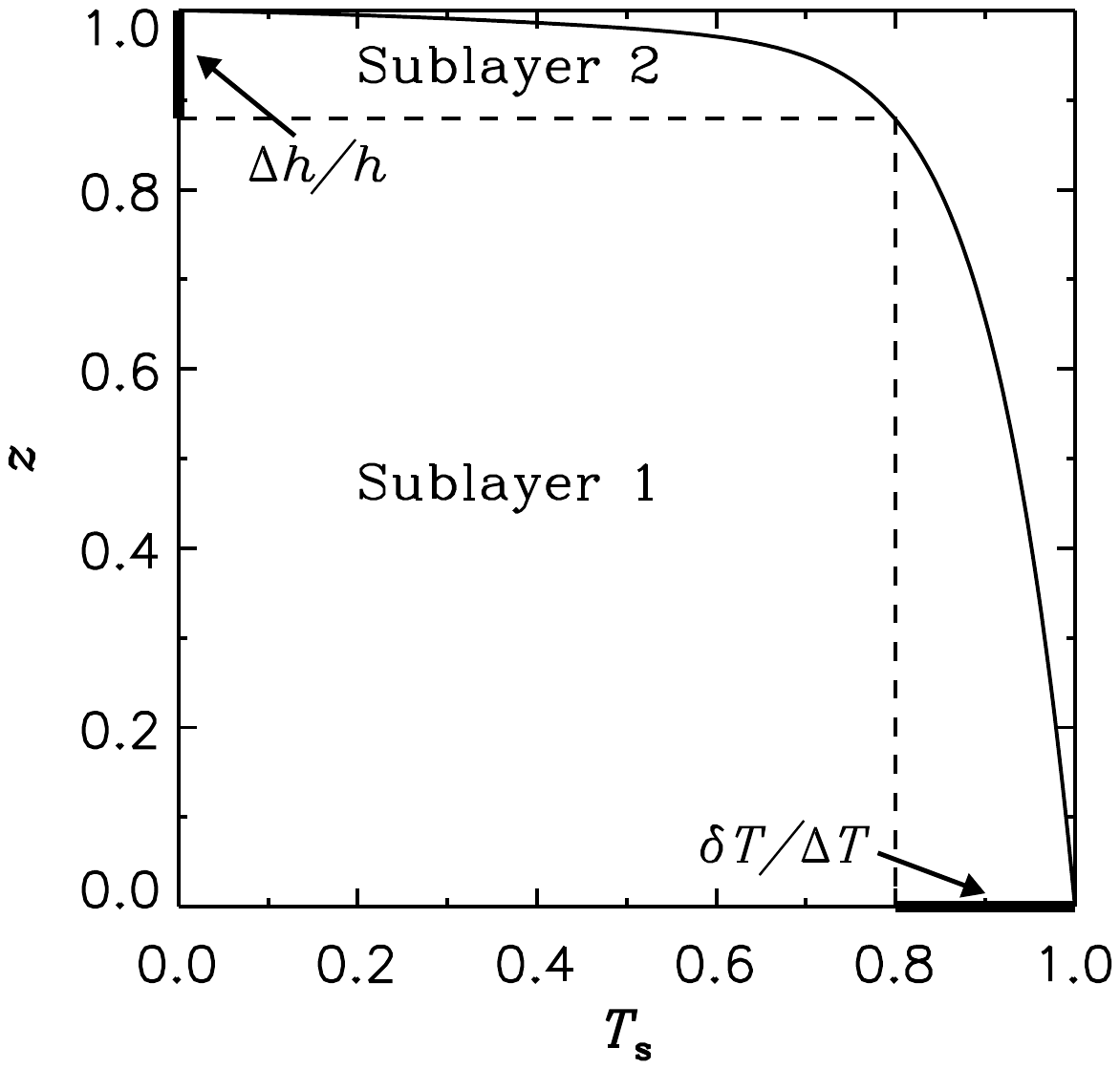}}\\
\tiny{(a)\hspace{6cm}(b)}
\caption{(a) Static $\chi(z)$ distribution and (b) static temperature profile $T_\mathrm s (z)$ corresponding to the law (\ref{chivsT}).}
\label{profile}
\end{center}
\end{figure}

We choose the temperature dependence of thermal diffusivity in the form
\begin{equation}
\label{chivsT}
\chi(T) = 1 + 5T+ 600T^{10},
\end{equation}
{with $\chi(T_\mathrm{top})/\chi(T_\mathrm{bot})=1/606$; from here on, we use $\Delta T$ as the unit temperature and the layer thickness $h$ as the unit length. In these dimensionless variables, the distributions of $\chi(z)$ and $T(z)$ for a motionless fluid due to the law (\ref{chivsT}) are shown in Figs~\ref{profile}a and \ref{profile}b, respectively.} The static temperature $T_\mathrm s$ varies little (by a dimensional quantity $\delta T \ll \Delta
T$) across the main portion of the layer (Sublayer~1) of a dimensional thickness $h - \Delta h$, where $\Delta h \ll h$, while the most part of the temperature difference,
$\Delta T - \delta T$, corresponds to Sublayer~2 with a small thickness $\Delta h$, near the upper surface.  The kink near $z=h - \Delta
h$ in the temperature profile specified in this way qualitatively resembles {(altough does not reproduce)} the transition from the bulk of the solar convection zone, where the stratification
is weekly superadiabatic, to the overlying strongly unstable layer with a depth of order 1~Mm.

We use an extended Boussinesq approximation, which admits thermal-diffusivity
variations \citep[see, e.g.,][for a discussion of different versions of this
approximation]{Getling:1998}. If, in addition to the above-specified variables, we choose the characteristic time of
viscous momentum transport $\tau_\nu = h^2/\nu$ as the unit time ($\nu$ being the kinematic viscosity), the governing equations assume the following
dimensionless form:
\begin{eqnarray}
&&\pder{\vec{v}}{t}
\,+\,({\vec{v}}{\vec{\cdot}}{\vec{\nabla}})\,{\vec{v}}\,=
\,-\,{\vec{\nabla}}\,{\varpi}\,+\,\hat{\vec{z}}\frac {\mathrm{Ra}}{\mathrm{Pr}}\,(T-T_\mathrm s)\,+\,\nabla^2{\vec v},\\
&&\pder{T}{t}\,+\,{\vec{v}}{\vec{\cdot}}{\vec{\nabla}}\,T\,=
\,\frac 1{\mathrm{Pr}} {\vec{\nabla}}\,{\vec{\cdot}}\,\frac{\chi(T)}{\chi(T_\mathrm{top})}{\vec{\nabla}}\,T,\\
&&{\vec{\nabla}}{\vec{\cdot}}{\vec{v}}\,=\,0.
\label{eqns}
\end{eqnarray}
Here, $t$ is the time,  $x,\,y,\,z$ are Cartesian coordinates, $\vec v$ is the
velocity vector, $\varpi$ is the pressure, $T$ is the temperature, $T_\mathrm s
(z)$ is the static temperature distribution, $\hat{\vec{z}}=(0,\,0,\,1)$ and
$$
\mathrm{Ra}=\frac{\alpha g\Delta T h^3}{\nu\chi(T_\mathrm{top})}\quad {\mathrm{and}}
\quad \mathrm{Pr}=\frac{\nu}{\chi(T_\mathrm{top})}
$$
are the Rayleigh and Prandtl numbers, $\alpha$ being the volumetric coefficient of thermal expansion of the fluid and $g$ the gravitational acceleration.
{The boundary conditions can be written in the following form:
\begin{equation}
\vec v|_\mathrm{bot}=\vec v|_\mathrm{top}=\vec v|_\mathrm{side}=0,\quad \quad   T|_\mathrm{bot}=1,\quad T|_\mathrm{top}=0,\quad \left.\pder{T}{\vec n}\right|_\mathrm{side}=0,
\end{equation}
where the subscript ``side'' refers to the side boundaries of the box and $\vec n$ is the vector normal to a side boundary.}

To solve the Navier--Stokes equations, we use the known
{\textit{Semi-Implicit Method for Pressure-Linked Equations} \citep[\mbox{SIMPLE}, see][]{Fletcher}} of a predictor--corrector class based on staggered grids, {modified by \citet{KMPdu,KMPmm}}.
In essence, it replaces the incompressibility equation with a Poisson equation
for pressure. First, a predictor for velocity is calculated from the equation
of motion at a current time layer. Next, a pressure corrector is found and used to correct the
velocity field so as to ensure incompressibity, {after which the temperature field is determined from the thermal-conduction equation}. The spatial approximation of
the equations is chosen based on the conservativeness requirement; it has a
second-order approximation inside the domain and a first-order approximation at the boundary. {An implicit conservative finite-difference scheme is constructed using a finite-volume method. Our computational grid is uniform and consists of $256\times 256\times 31$ points. (We also used finer grids in tentative runs; they took much more time but gave very similar results. The ultimately chosen grid offers a reasonable compromise between the computational-time consumption and the accuracy achieved.) The time step of computations is $\tau=10^{-2}\tau_\nu$.}

Initially, the fluid is motionless. The flow originates from random
temperature perturbations; {to reduce the time needed to achieve a quasi-steady state, we introduce them at a certain height inside Sublayer~2, which is convectively most unstable because of the high static-temperature gradient}.

\section{Simulation results}

\subsection{Qualitative description}

\begin{figure}
\vspace{-4.2cm}
\begin{center}
\includegraphics[width=0.48\textwidth, bb=30 25 375 340,clip]{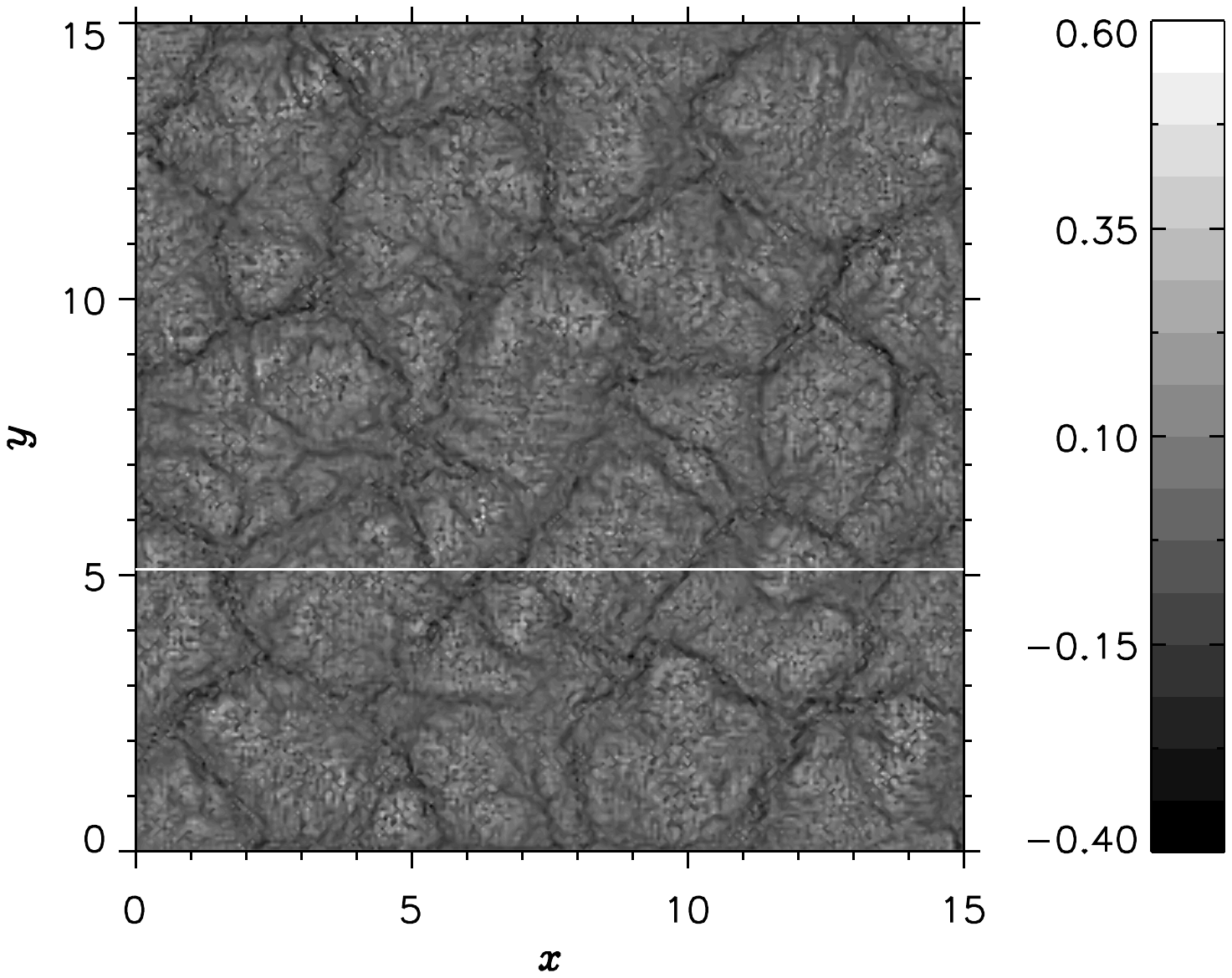}
\includegraphics[width=0.48\textwidth, bb=30 25 375 340,clip]{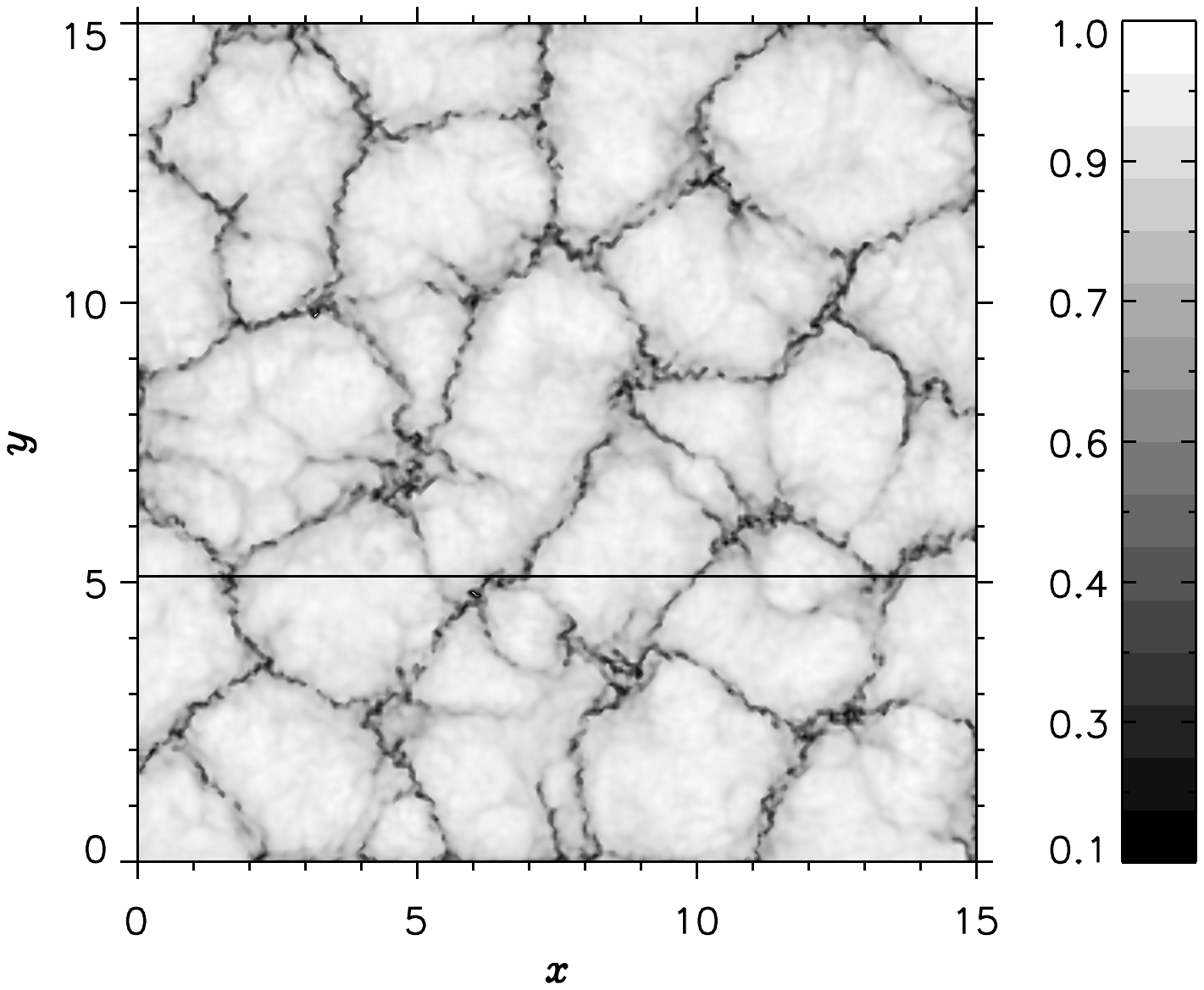}\\
\hspace{0.7cm}\tiny{(a)\hspace{6.0cm}(d)}\\[0.2cm]
\hspace{6.65cm}\includegraphics[width=0.475\textwidth,bb=50 35 818 215, clip] {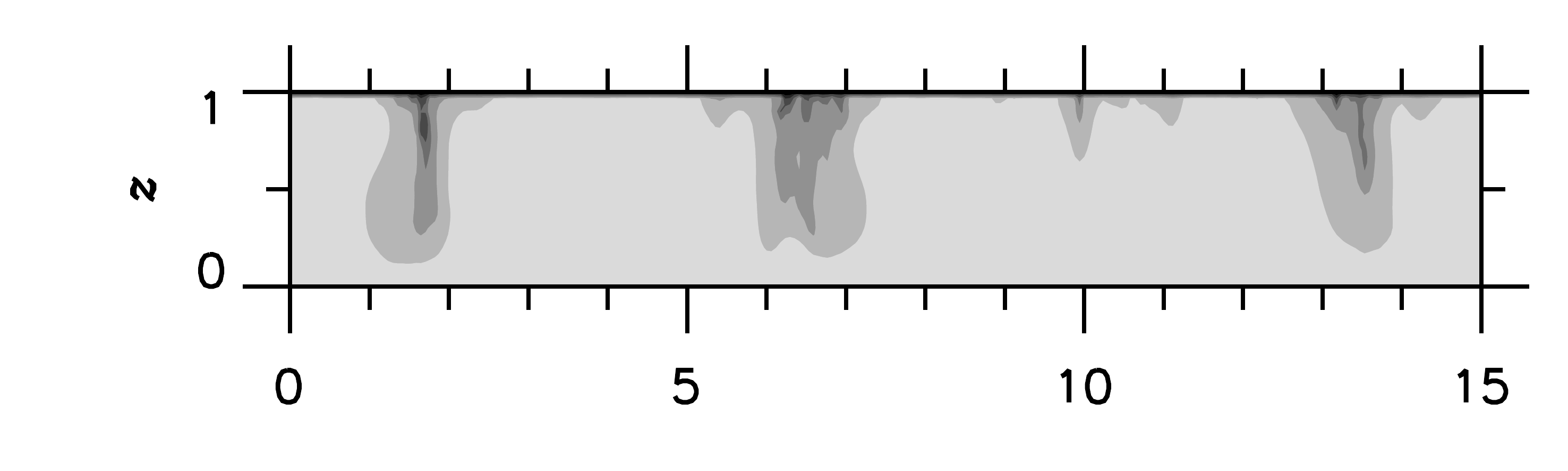}\\
\hspace{7.05cm}\tiny{(e)}\\[0.1cm]
\includegraphics[width=0.48\textwidth, bb=30 25 375 340,clip]{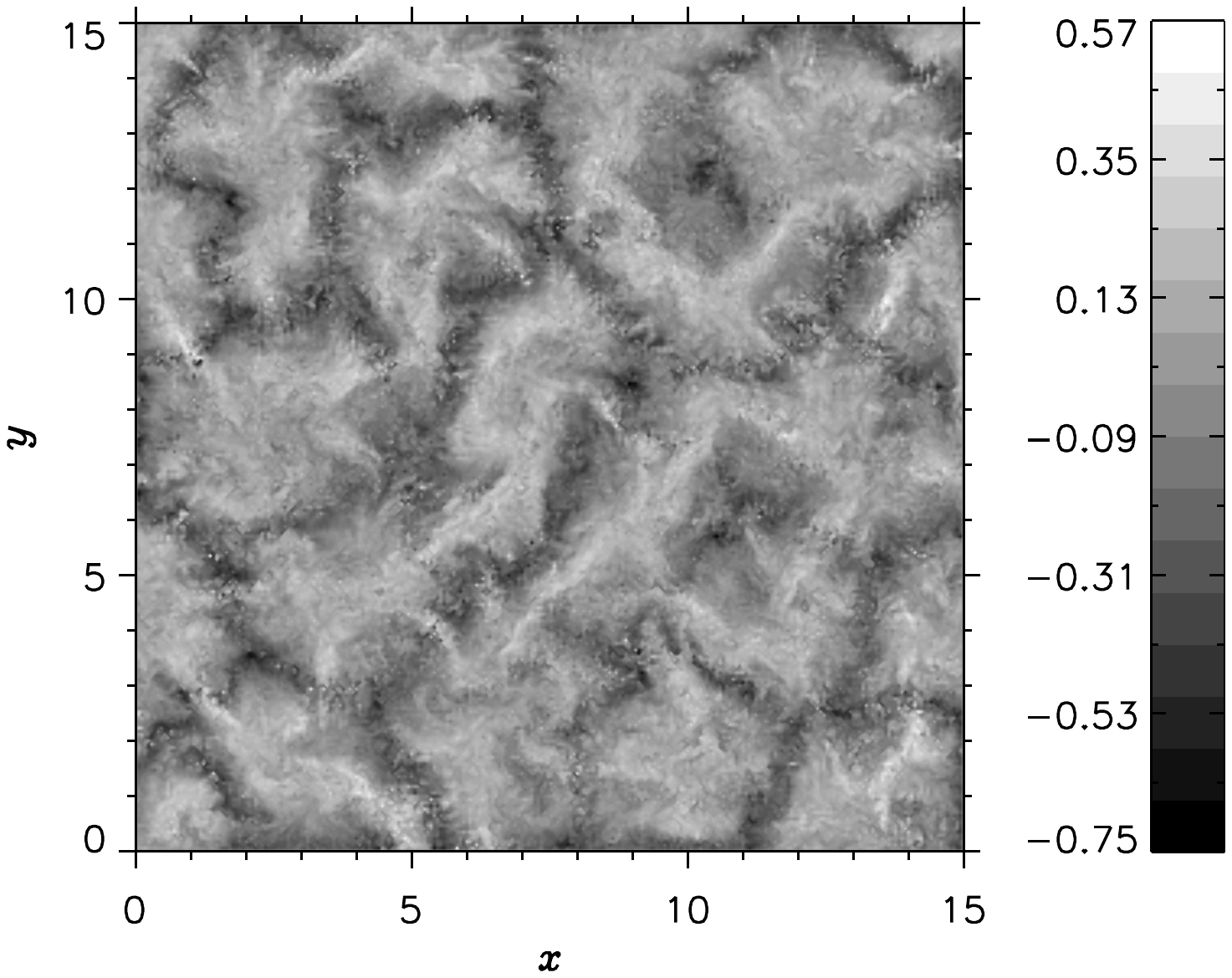}
\includegraphics[width=0.48\textwidth, bb=30 25 375 340,clip] {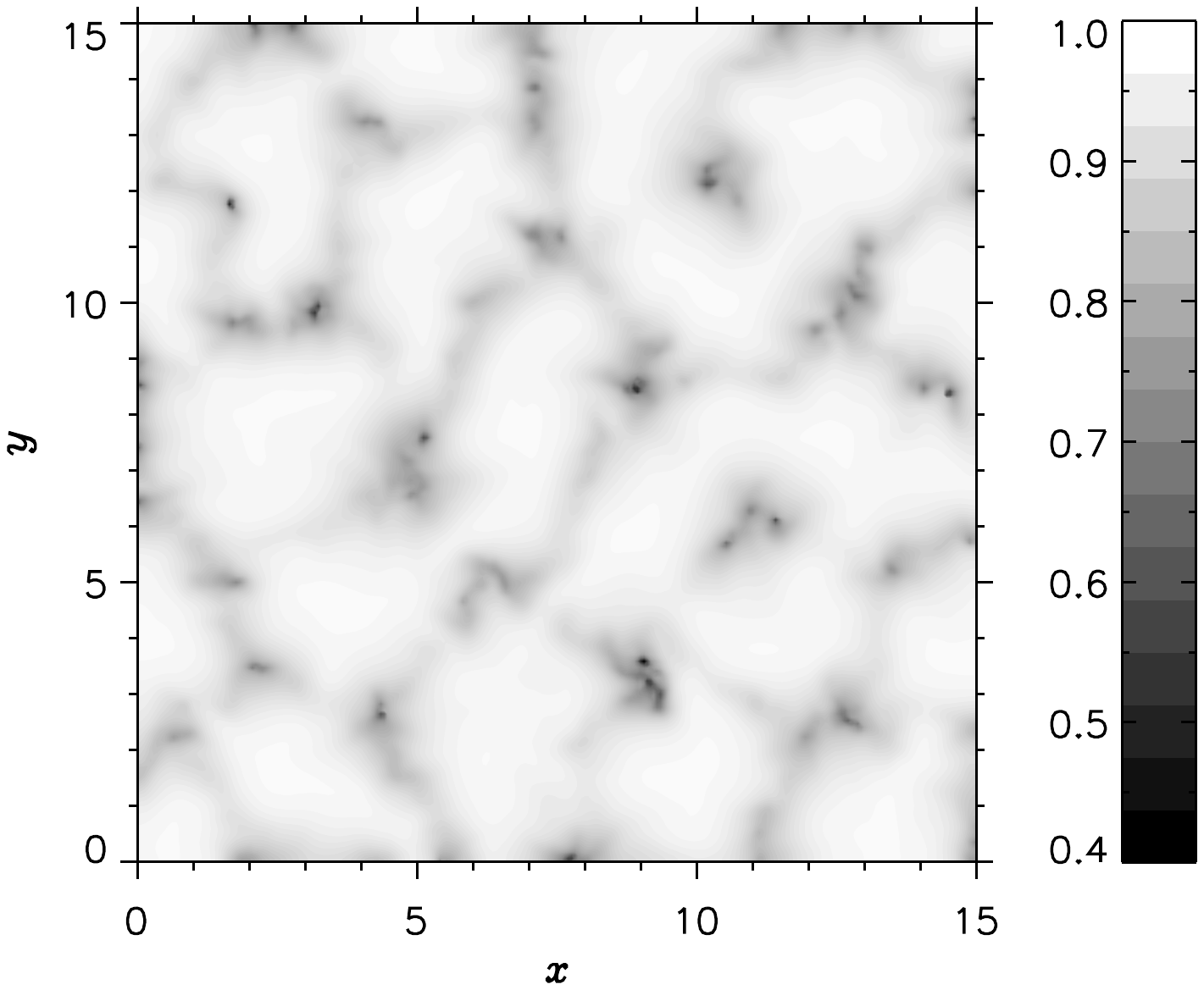}\\
\hspace{0.7cm}\tiny{(b)\hspace{6.0cm}(f)}\\[0.1cm]
\includegraphics[width=0.48\textwidth, bb=30 10 375 340,clip]{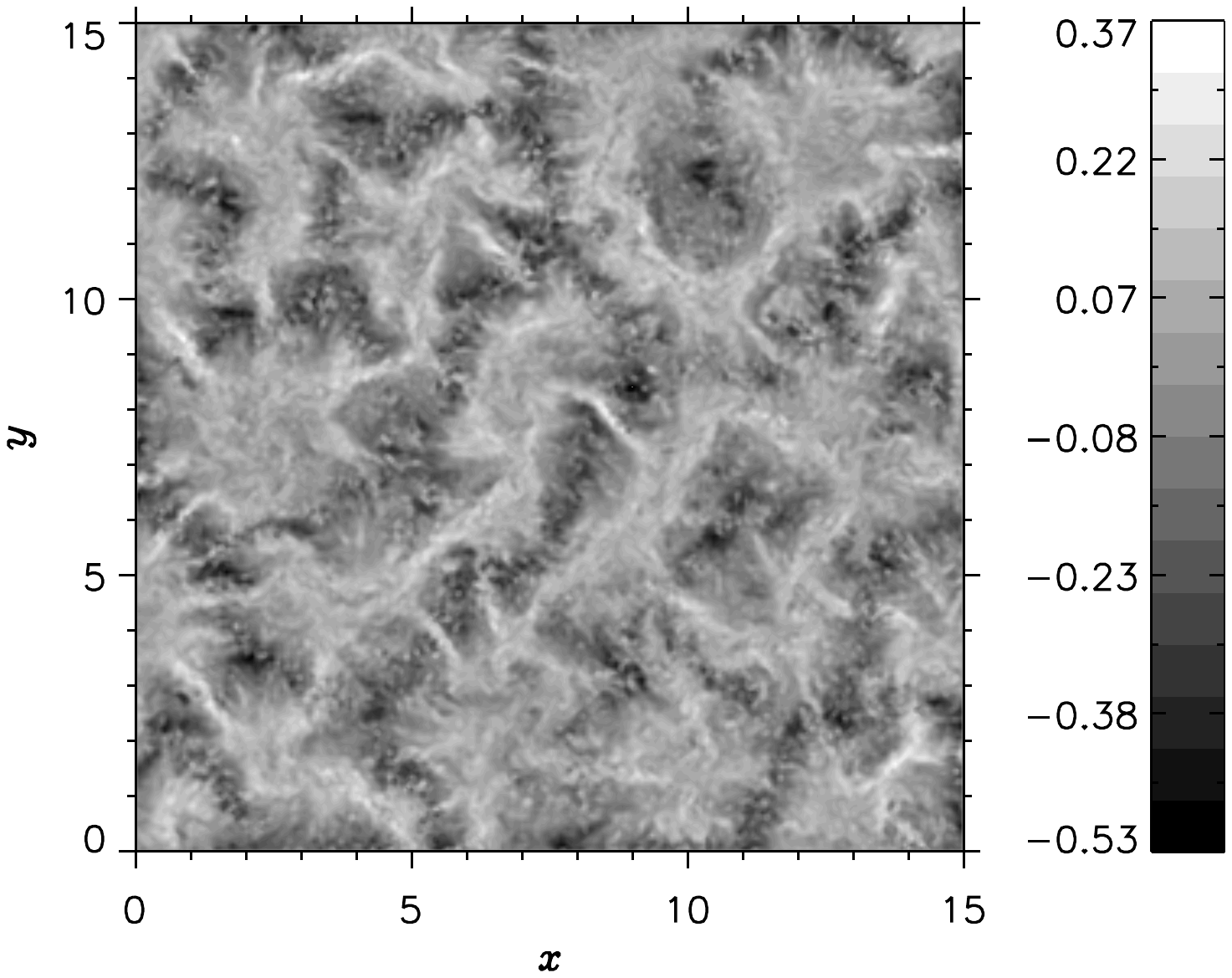}
\includegraphics[width=0.48\textwidth, bb=30 10 375 340,clip] {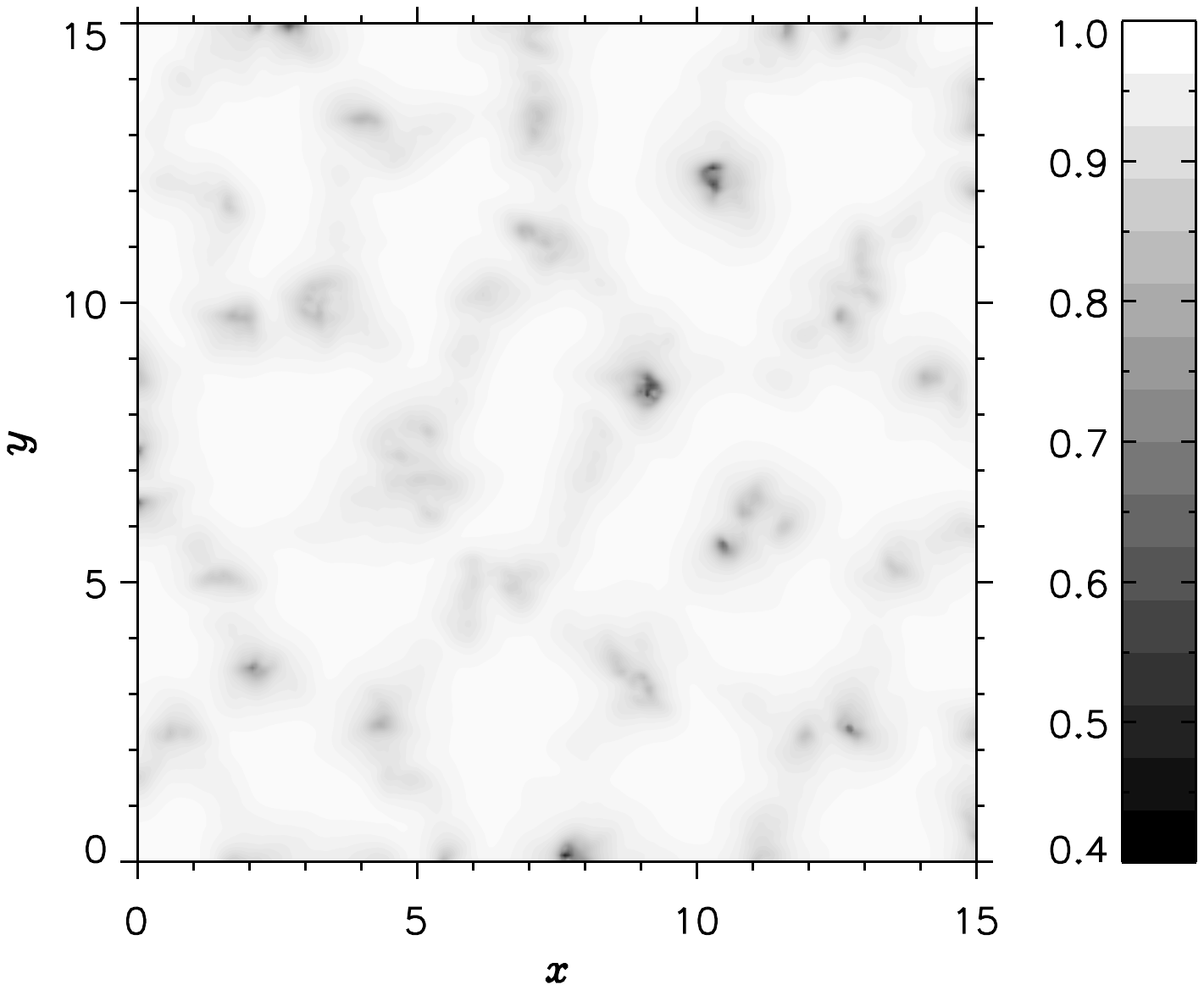}\\
\hspace{0.7cm}\tiny{(c)\hspace{6.0cm}(g)}\vspace{-0.2cm}
\caption{{Flow pattern at time $t=170\tau_\nu$:} (a)--(c) Vertical velocity component at $z=0.97, 0.323, 0.161$, with variation ranges
[$-0.44, 0,64$], [$-0.75, 0.60$], [$-0.80, 0,49$], respectively;
(d), (f), (g) temperature field at the same levels with variation ranges [0.01, 0.99], [0.45, 0.99],
[0.49, 1.00], respectively; (e) temperature distribution in the vertical section $y=5.11$ marked in (a), (b) with horizontal lines (the $z$-scale is exaggerated with respect to the $x$-scale
to show up the structure of the downdrafts).}
\label{struct}
\end{center}
\end{figure}

We consider here a representative computational run for which $\mathrm{Ra}=1.5\times 10^8 \approx 37.5\,\mathrm{Ra}_\mathrm c$ and $\mathrm{Pr}=1$, where the critical Rayleigh number is $\mathrm{Ra}_\mathrm c \approx 4 \times 10^6$ {(it was determined in the course of simulations as the $\mathrm{Ra}$ value at which the temperature profile begins departing from its static shape)}. {The run lasted until $t=170\tau_\nu$, or $17000\tau$.} The flow starts developing in Sublayer~2 and gradually involves the whole layer in motion. {By time $t\approx 80\tau_\nu$, it reaches a quasi-steady state of full development, with slow variations in the structure of the flow and with its almost constant kinetic energy.}

\begin{figure}[!t]
\begin{center}
\includegraphics[width=4.4cm,bb=40 0 420 360pt, clip]{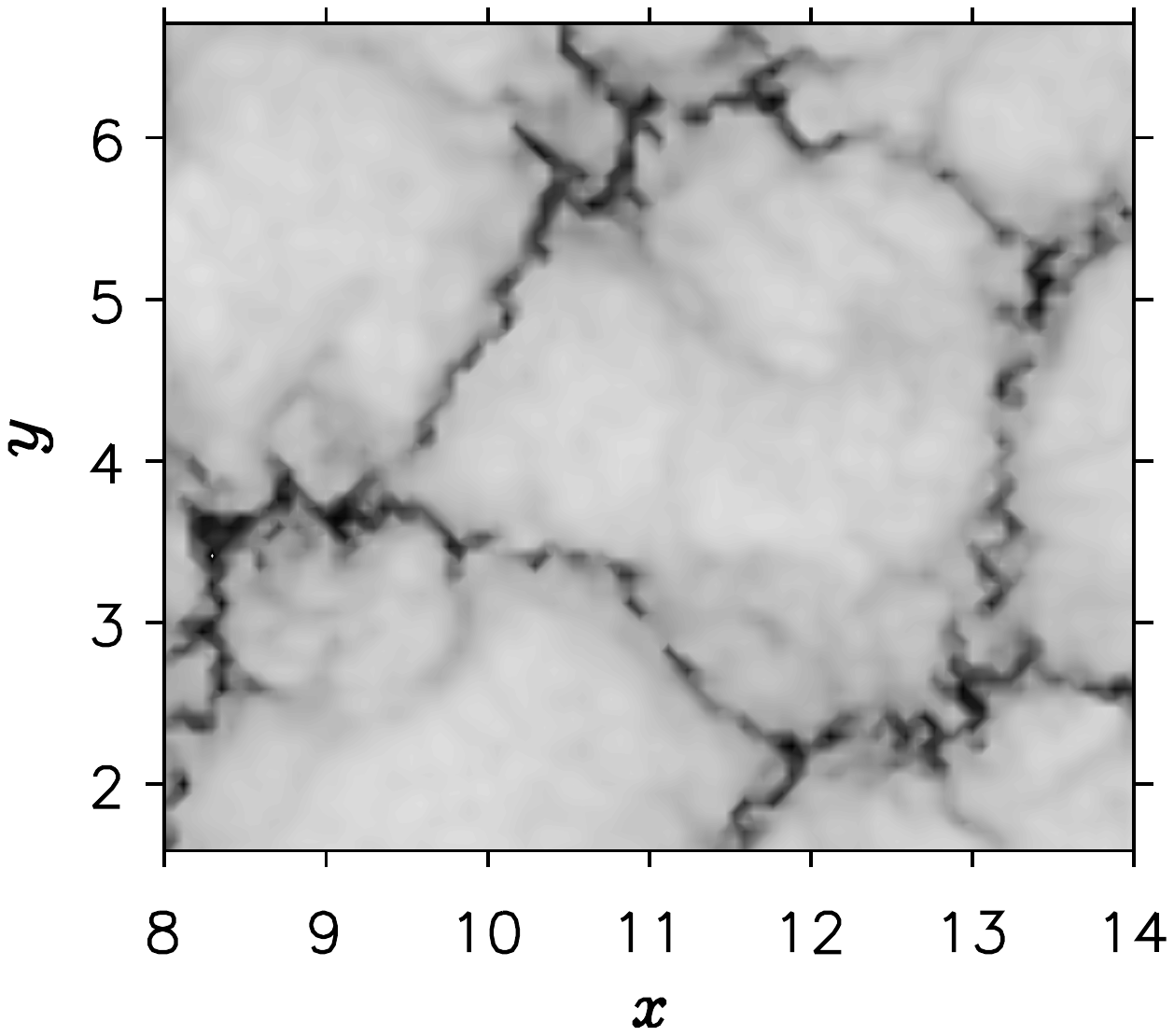}
\includegraphics[width=4.4cm,bb=40 0 420 360pt, clip]{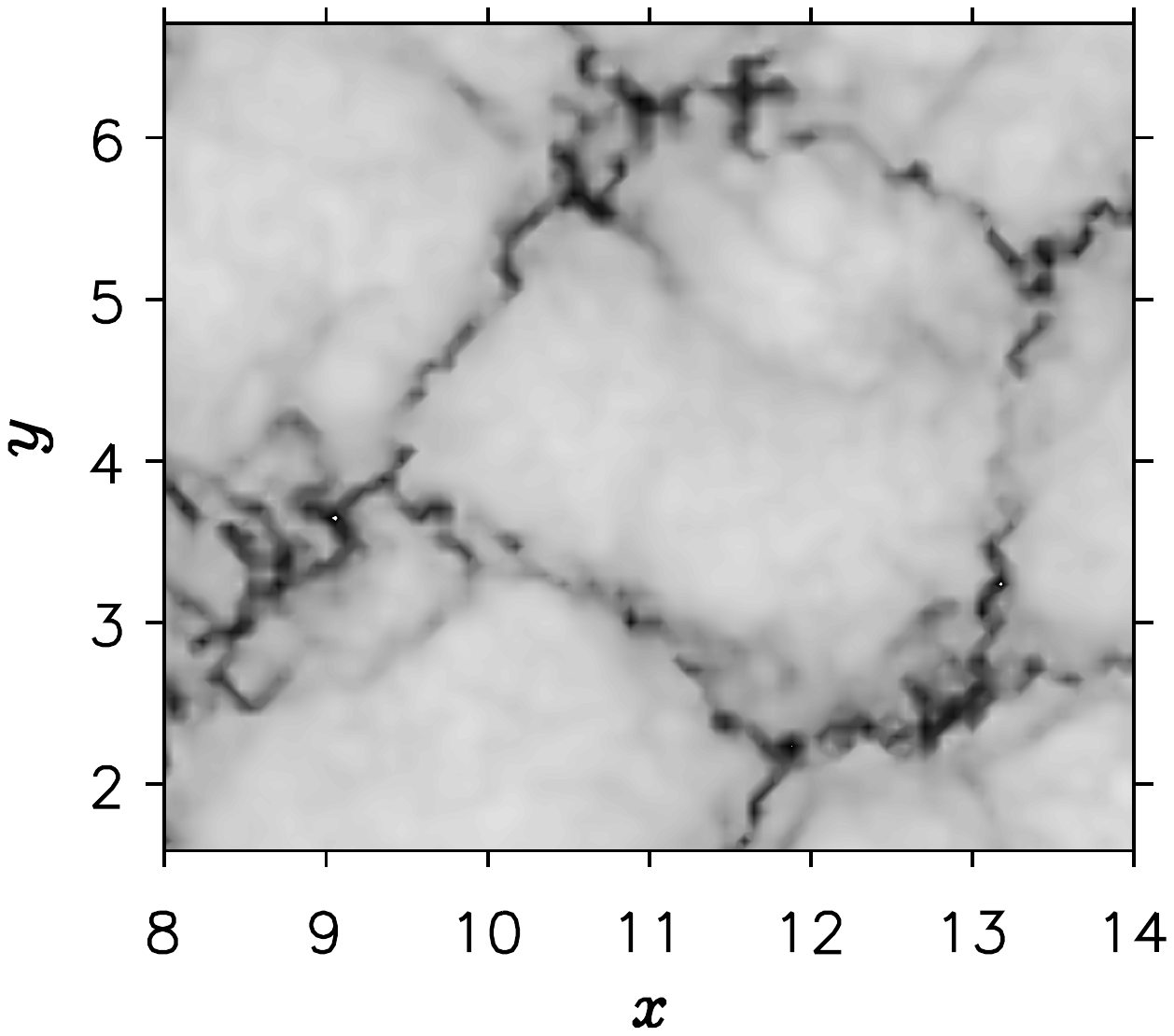}
\includegraphics[width=4.4cm,bb=40 0 420 360pt, clip]{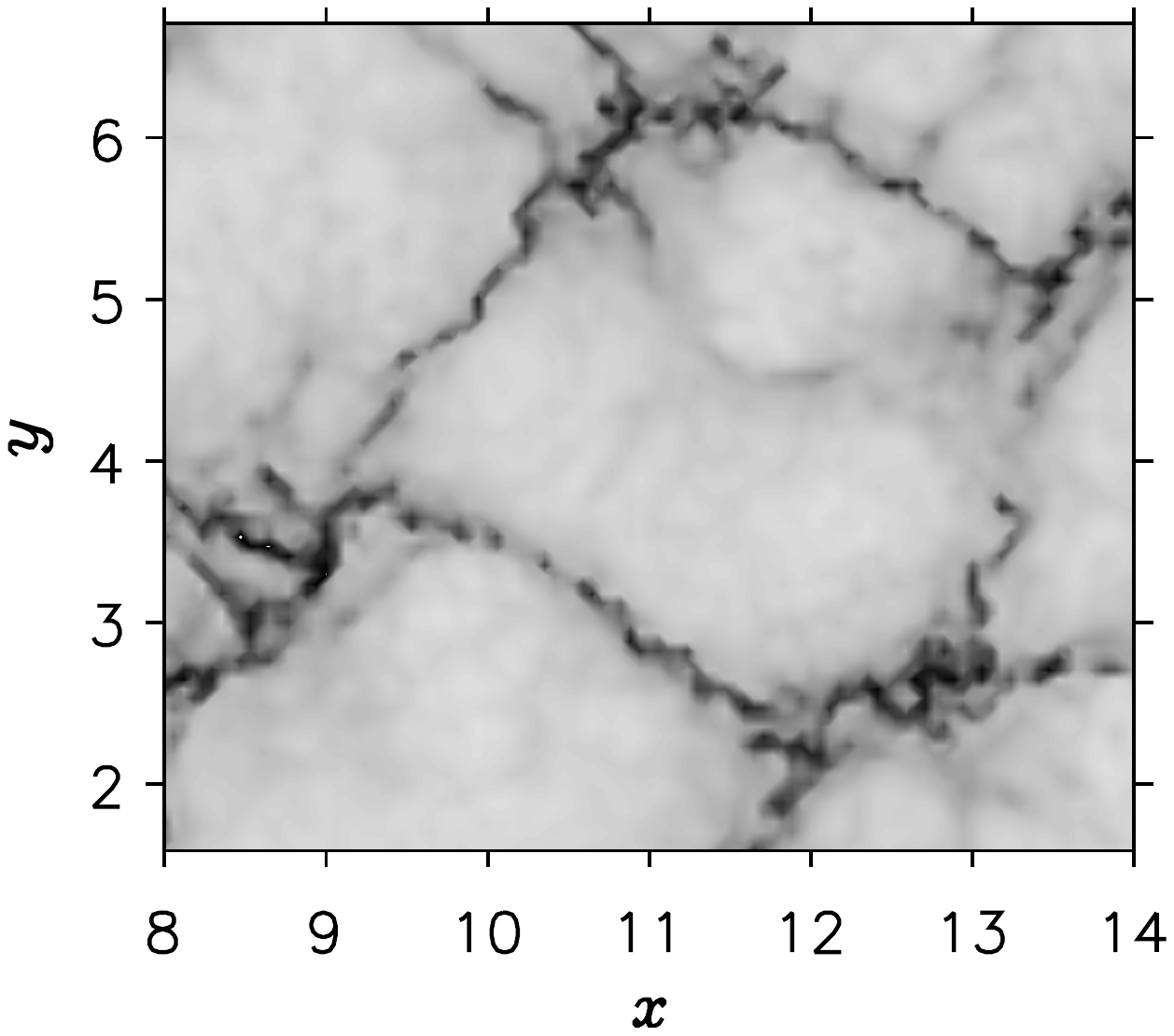}\\
\caption{From left to right: an enlarged fragment of the temperature
field at $z=0.97$ and times $t=134.72$, 136.35, 137.99 (measured in units of $\tau_\nu$). The largest, first-scale cells (outlined by downflow lanes, which appear dark grey) are divided by light-grey isthmuses into smaller, second-scale structures. The smallest, third-scale features manifest themselves most clearly in dark mottles near the first-scale intercellular lanes, especially at its nodes. The smallest features are advected by the flows
inside the larger, first-scale and second-scale cells.}
\label{motions}
\end{center}
\end{figure}

Figures \ref{struct}a--\ref{struct}c show the distributions of the vertical velocity component over horizontal sections of the computational domain at three levels {and time $t=170\tau_\nu$, and Figs~\ref{struct}d, \ref{struct}f, \ref{struct}g represent the temperature distributions in the same sections.} The large cellular structures (especially clearly visible in Fig.~\ref{struct}d) exhibit, during some time interval, a tendency of increasing their sizes, with slight deformations and drift. They gradually fill the entire depth of the layer, and their growth virtually
terminates by times $t \approx 65 \tau_\nu$. We designate these structures as the \emph{first-scale} cells.
In the well-developed convection pattern, a multitude of small features are observed at levels slightly below the upper layer boundary. They move following their own laws and approaching the borders of the large structures.
These \emph{third-scale} cells emerge over the entire area of each first-scale cell. Cells of an intermediate, \emph{second scale} are also present; {they can be seen in enlarged images of the temperature field (Fig.~\ref{motions})}. As we shall demonstrate below (in Sections~\ref{small} and \ref{spectral}), this scale hierarchy of structures can be made most distinct by applying smoothing and spectral-processing techniques.

The temperature variations over the horizontal section $z=0.97$ (see Fig.~\ref{struct}d) are within the range $T_{\max} -T_{\min} =
0.98$. The borders of the first-scale structures coincide in their location with the strongest downdrafts. {Similar cold downdrafts related to the structures of smaller scales reach smaller depths (see Fig.~\ref{struct}e), and the usual order-of-magnitude agreement between the horizontal and vertical sizes of a convection cell holds. This clearly indicates that the second-scale and third-scale cells are ``suspended'' near the top layer boundary.} The flow near the bottom surface is not so diverse: there are only several isolated hot ascending plumes, which are relatively far apart.
\begin{figure}
\begin{center}
\includegraphics[width=0.47\textwidth,bb=40 6 384 340,clip]{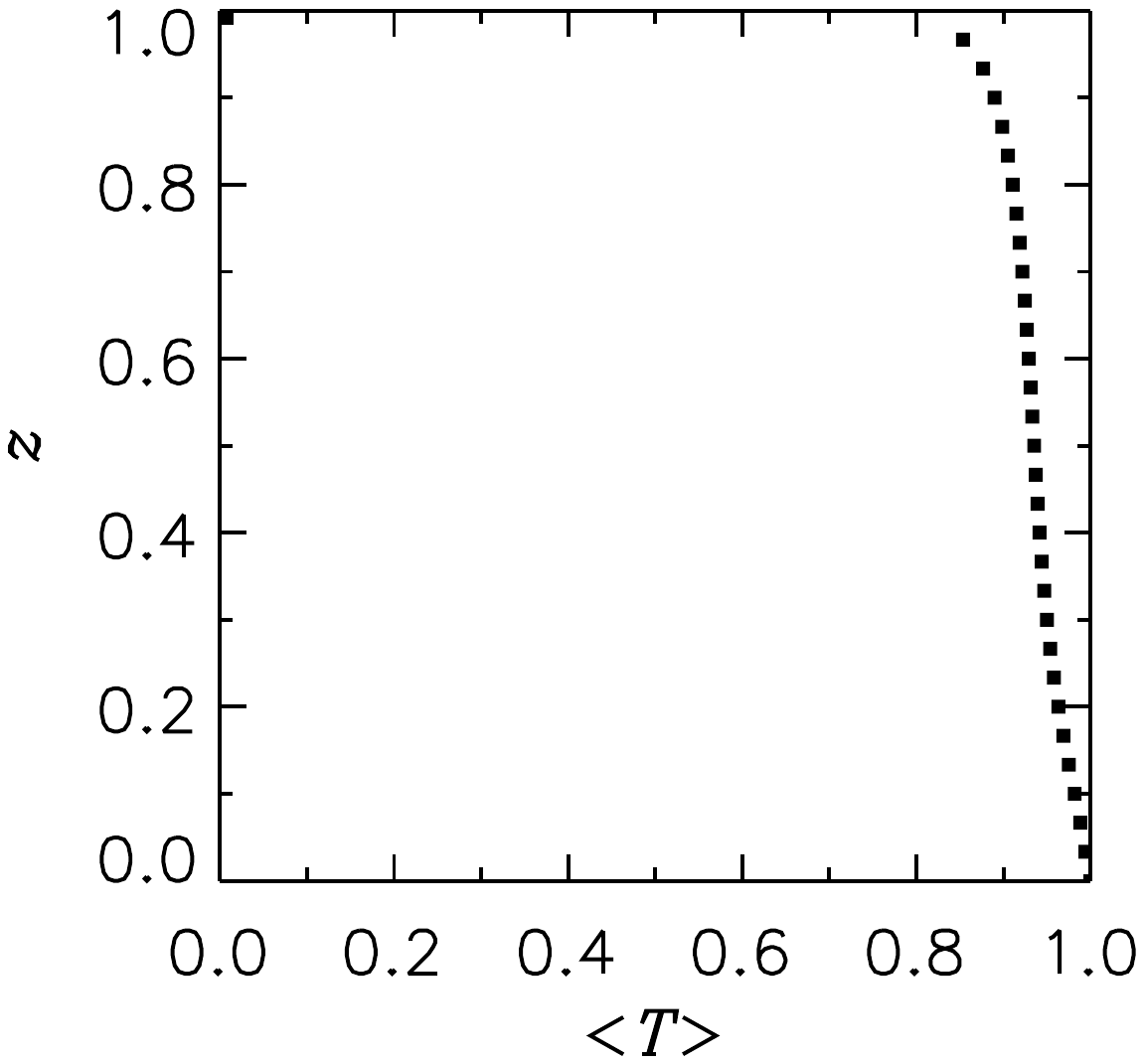}
\includegraphics[width=0.45\textwidth,bb=40 0 345 300,clip]{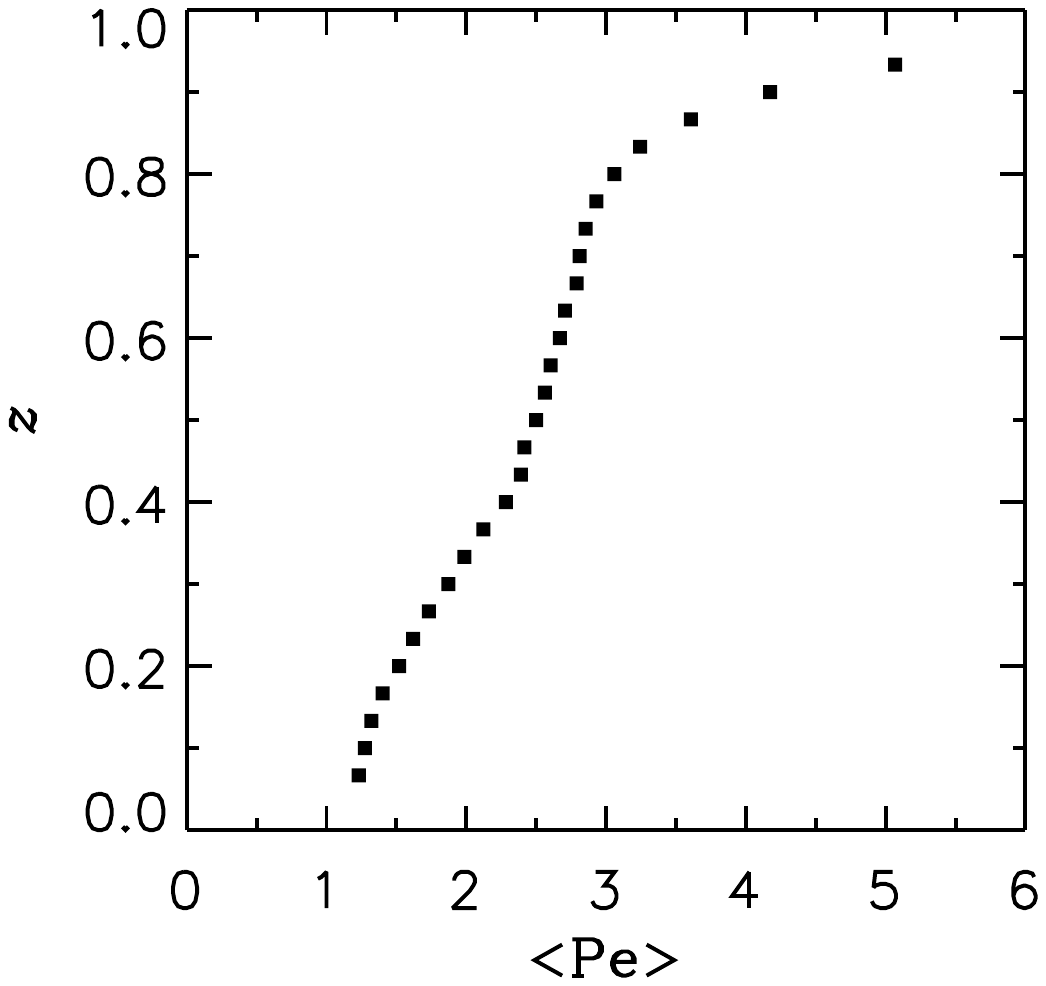}
\caption{{Height variation of the temperature and P\'eclet number averaged over the horizontal plane.}}
\label{Pevsz}
\end{center}
\end{figure}

{The relative, advective-to-diffusive, efficiency of heat transport can be judged by the P\'eclet number,
\begin{equation}\label{Pe}
    \mathrm{Pe}=\frac{VL}{\chi};
\end{equation}
here, $V$ is the characteristic velocity and $L$ is the
characteristic spatial scale of velocity variation in the directions normal to
the velocity vector. To estimate the \emph{local} P\'eclet number, we assume
that
$$V=\max\{ |v_x|,|v_y|,|v_z| \}$$ and estimate the velocity shear as
$$s=\max\left\{\left|\pder{v_x}{y}\right|,\,\left|\pder{v_x}{z}\right|,\,\left|\pder{v_y}{x}\right|,
\,\left|\pder{v_y}{z}\right|,\,\left|\pder{v_z}{x}\right|,\,\left|\pder{v_z}{y}\right|\right\};$$
then $$L\sim\frac{V}{s}$$ and
$$\mathrm{Pe}\sim\frac{V^2}{s\chi}.$$
The partial derivatives can be calculated in a standard way, using
their finite-difference analogues.}

{The height variations of $\langle\ T\rangle$ and
$\langle\mathrm {Pe}\rangle$ (where the angle brackets designate averaging the
local values over the horizontal section of the box) are plotted in
Fig.~\ref{Pevsz}. As can be seen, convective mixing at the chosen high Rayleigh number results in a relatively weak $z$-variation of temperature in the bulk of the layer; $\langle\mathrm {Pe}\rangle$ acquires its largest values approaching the top boundary.}

\begin{figure}
\begin{center}
\includegraphics[width=0.48\textwidth, bb=30 10 370 340, clip]{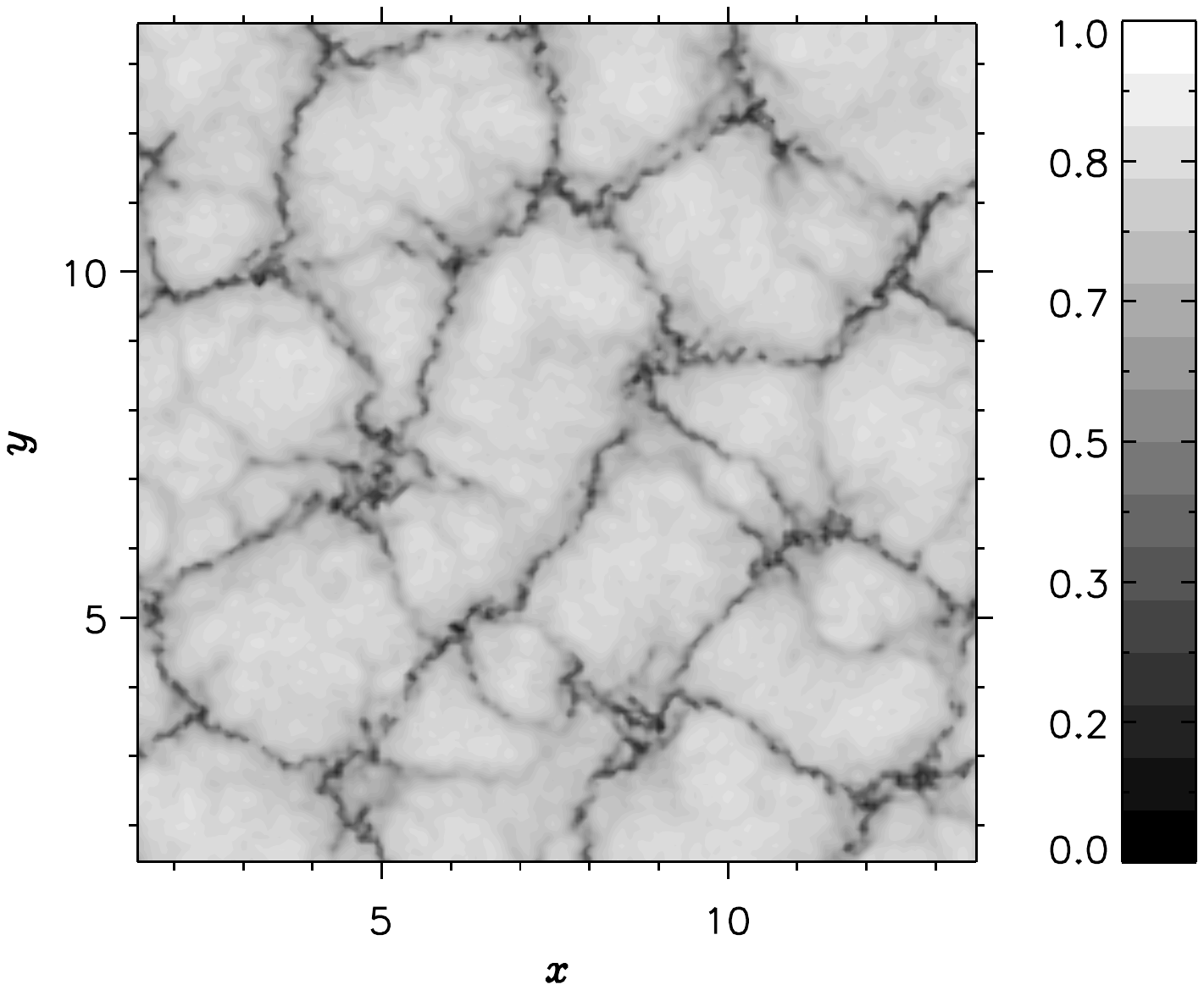}
\quad\includegraphics[width=0.48\textwidth, bb=30 10 370 340, clip]{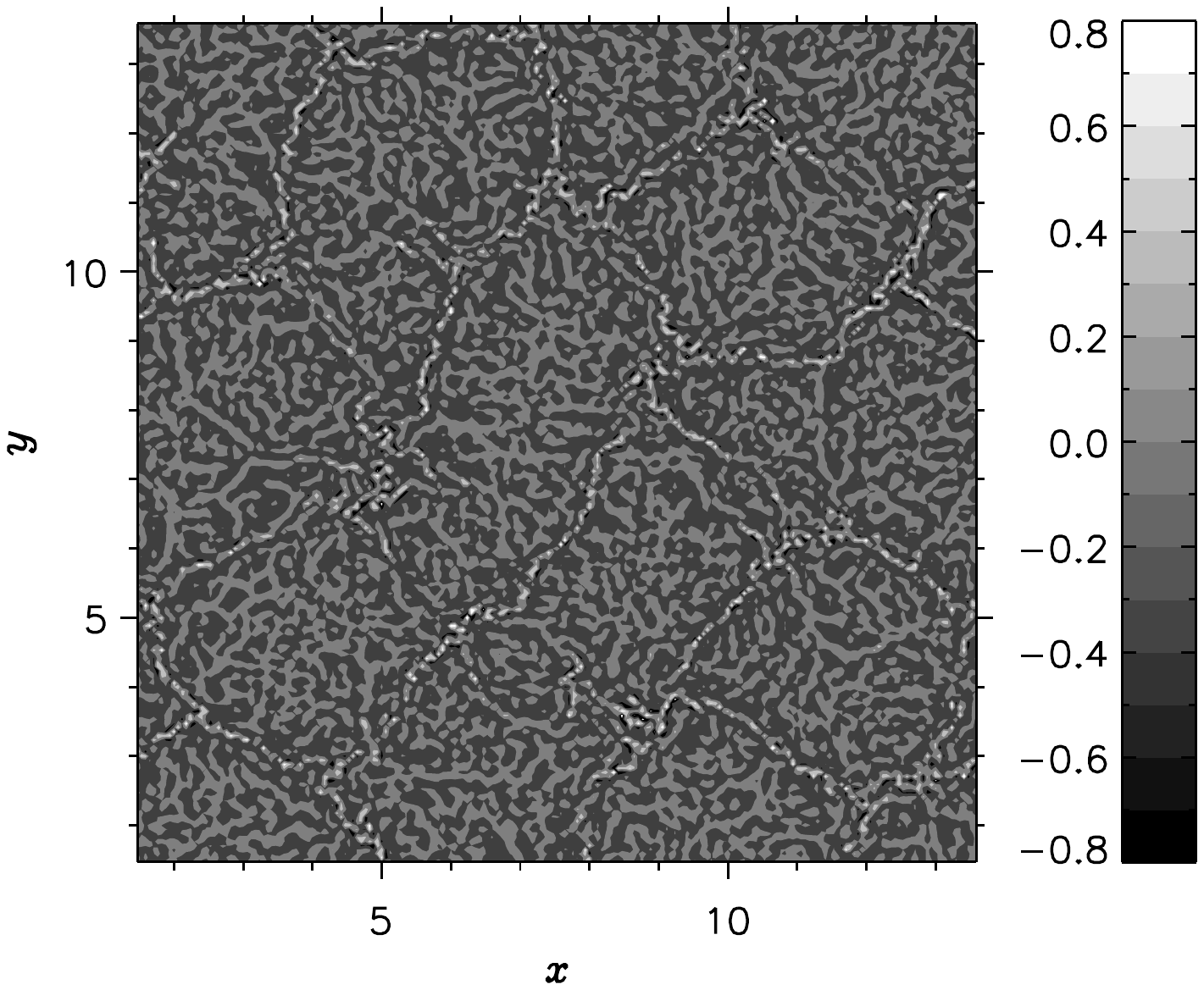}\\
\hspace{1cm}\tiny{(a)\hspace{6.8cm}(b)}
\caption{Processing of the temperature field at $z=0.97$ {(shown in Fig.~\ref{struct}d)} with a Gaussian (moving-average) filter:
(a)~long-wavelength component, averaging result (a variation range of [0.15,     1.0]); (b) short-wavelength component obtained by subtracting the average from the original field (a variation range of [$-1.2, 2.0$]).}
\label{moving}
\end{center}
\end{figure}

\subsection{Smoothing the field and singling out the small-scale
features}\label{small}

\begin{figure}[!t]
\begin{center}
\vspace{0.25cm}
\includegraphics[width=6cm, bb=20 0 370 350,clip]{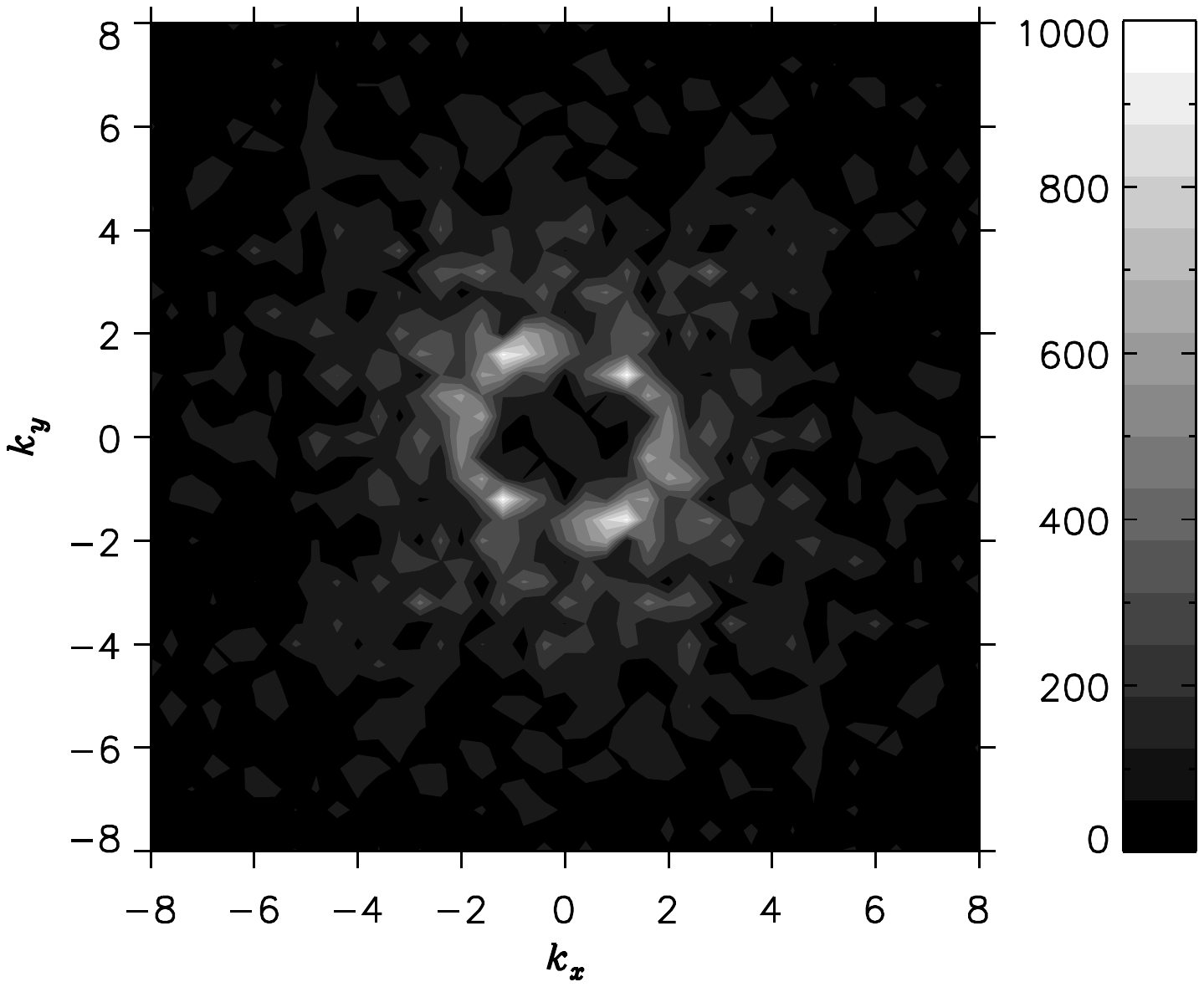}\hspace{1.2cm}
\includegraphics[width=6cm, bb=20 0 370 350,clip]{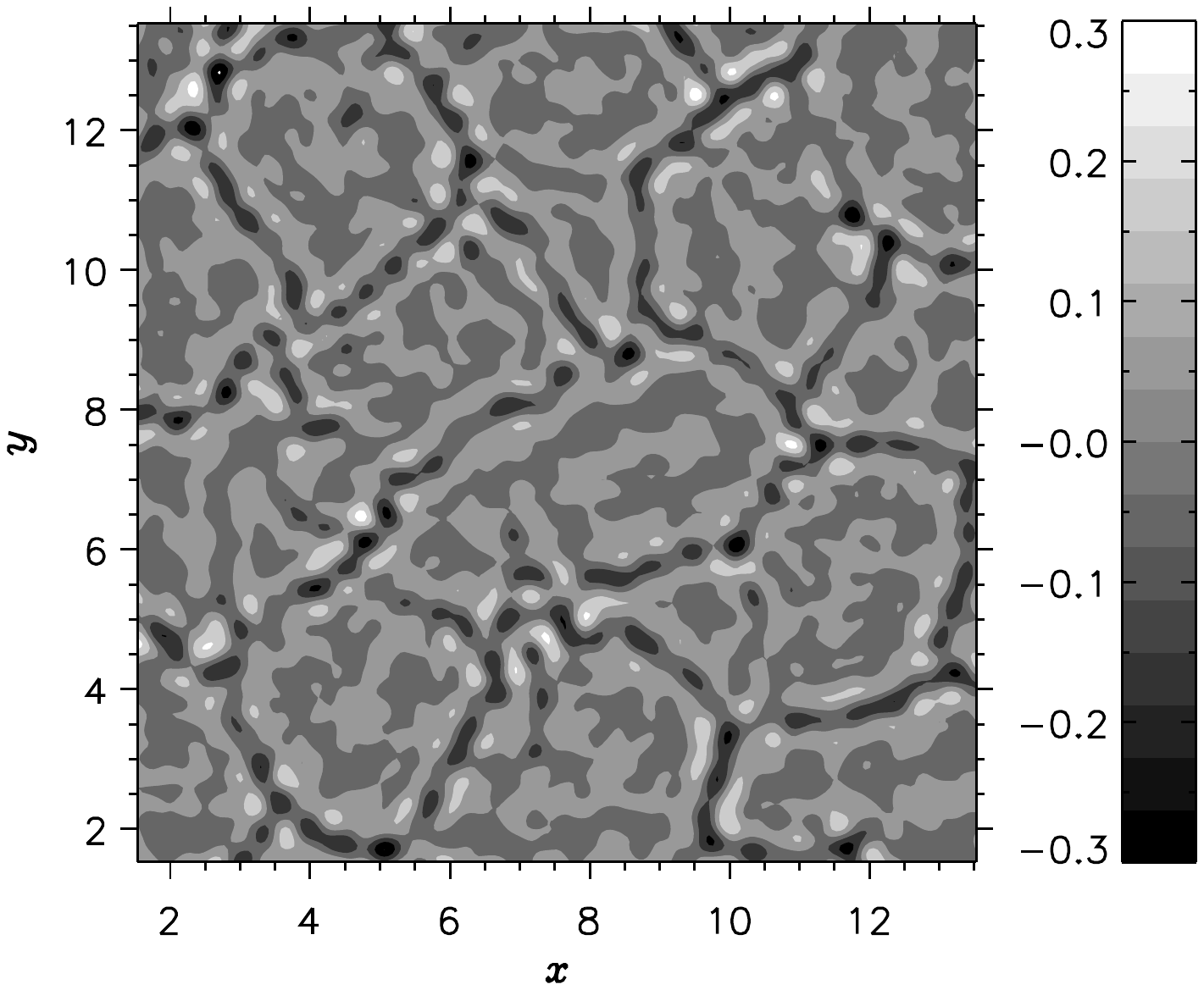}\\
\tiny{(a)\hspace{7.0cm}(b)}\\
\includegraphics[width=6cm, bb=20 0 370 350,clip]
{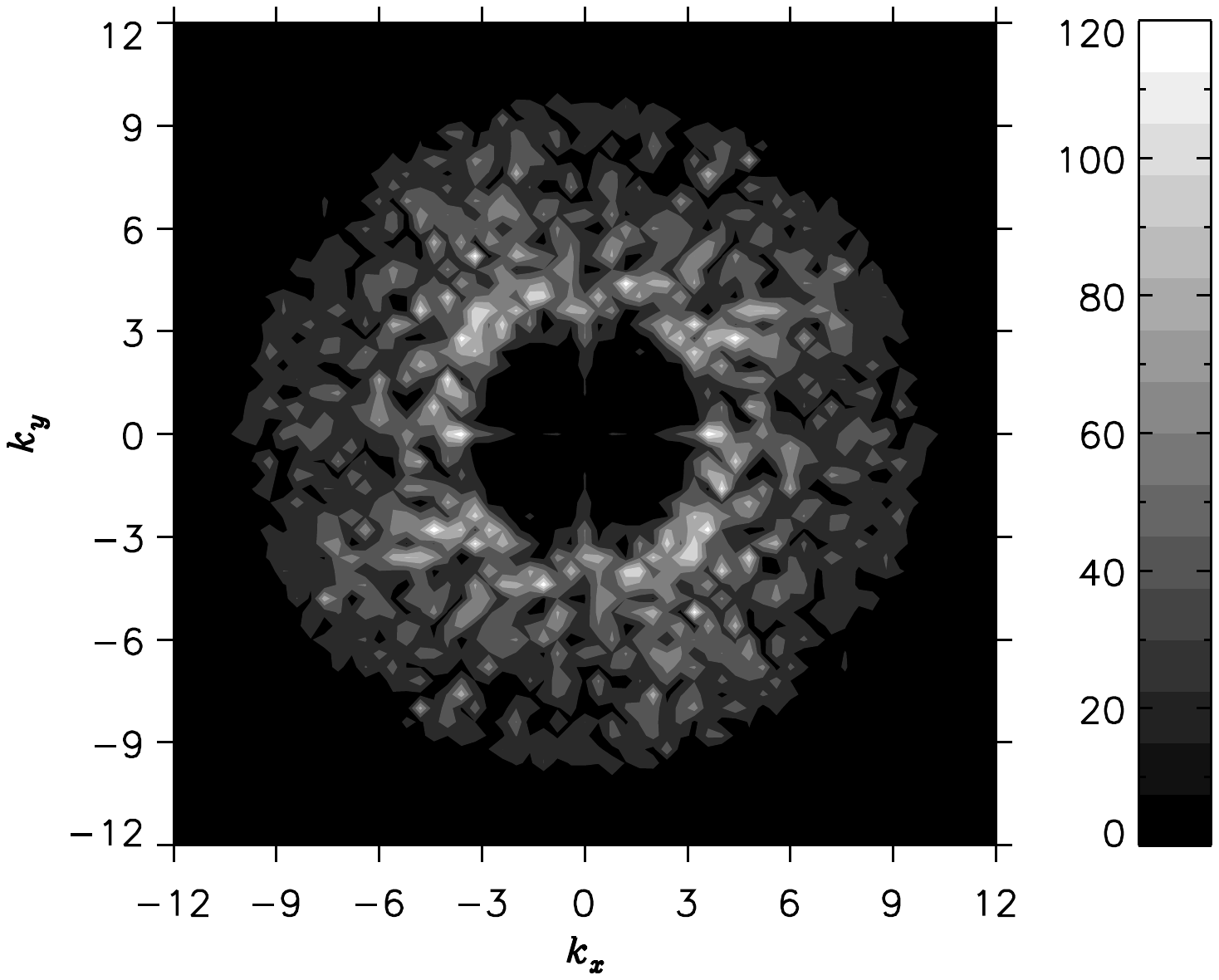}\hspace{1.2cm}
\includegraphics[width=6cm, bb=0 -20 445 340,clip]{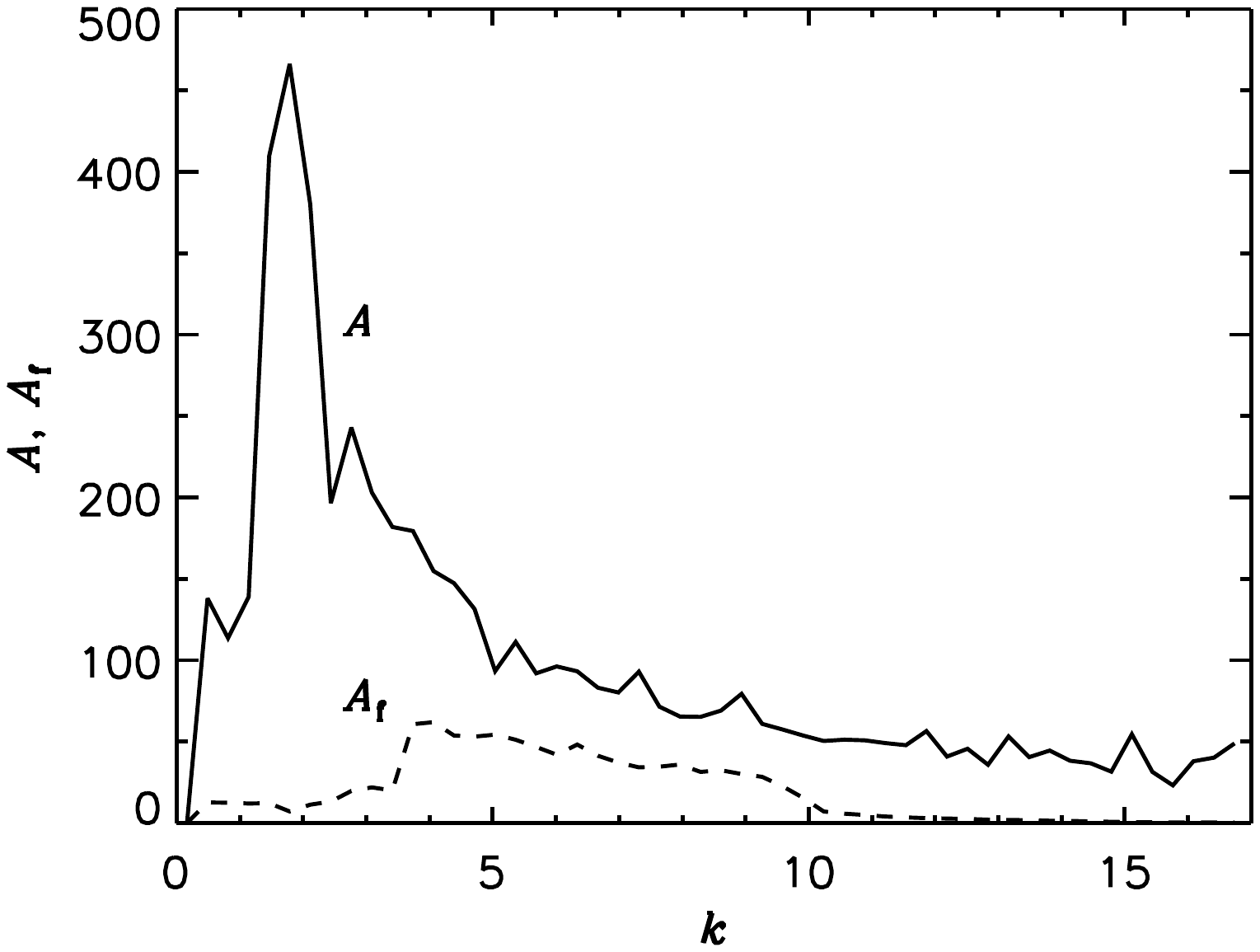}\\
\tiny{(c)\hspace{7.0cm}(d)}             
\caption{(a) Spatial Fourier-amplitude spectrum of the temperature field at $z=0.97$ (a variation range of [1,1050]); (b)~result of the processing of this temperature field with a band-pass filter (in physical space); (c)~spectrum of the processed field (b) (a variation range of [0,138]); {(d) spectra of the original and the processed field ($A$ and $A_\mathrm f$, respectively) averaged over all wavevectors with the same wavenumber $k=|\mathbf k|$.}}
\label{spectrum}
\end{center}
\end{figure}

To single out different flow scales, we apply a {standard} Gaussian moving-average filter to the temperature field. Figure~\ref{moving}a shows the averaging {(smoothing)} result; Fig.~\ref{moving}b, the result of subtracting the average from the original flow pattern. The averaged, long-wavelength component of the flow {(Fig.~\ref{moving}a) is visually very similar to the original shown in Fig.~\ref{struct}d but differs from it by the small-amplitude, short-wavelength component, Fig.~\ref{moving}b. The former} is basically represented by the largest (first-scale) structures separated by dark lanes; however, the
averaging procedure also reveals some ``bridges'', or ``isthmuses'' (appearing
slightly darker than the interiors of the first-scale cell on the whole), which were noticeable even in the original flow map. These ``bridges''
reflect the presence of structures of an intermediate, second scale.

The smallest (third-scale) structures (Fig.~\ref{moving}b) are ubiquitous in the whole horizontal section. As noted above, the periphery of the large (primary)
structures is formed by the strongest downdrafts, which run through {most part of the layer thickness, from top to deep levels; the downdrafts that permeat only a lesser part of the layer depth (Fig.~\ref{struct}e) are related to smaller-scale structures}.

\subsection{Spectral properties of the flow}\label{spectral}

Now, to comprehend how the flow spectrum reflects the presence of the three
convection scales revealed in our simulations, we analyse the flow structure
using spectral techniques. Figure \ref{spectrum}a represents the spatial spectrum of the original temperature field at $z=0.97$ (Fig.~\ref{struct}d) {in the form of a two-dimensional distribution of the Fourier amplitude \textit{versus} the wavevector $\mathbf k=\{k_x,k_y\}$, while Fig.~\ref{spectrum}d shows the $k$-variation of this amplitude averaged over all wavevectors $\mathbf k$ inside a narrow annular bin in the vicinity of a given wavenumber $|\mathbf k|=k$ (we denote this average as $A$). In the $(k_x,k_y)$ plane (Fig.~\ref{spectrum}a), there is a pronounced, slightly irregular spectral ring corresponding to the main peak of $A(k)$ in the range $1.2\lesssim k\lesssim 2.3$ (Fig.~\ref{spectrum}d); it represents the large (first-scale) cells visually identifiable in both the velocity and the temperature field (Figs~\ref{struct}a, \ref{struct}d).}

{To detect the small-scale structures filling the entire domain,
we apply a standard two-dimensional band-pass filter to the temperature pattern at $z=0.97$: in the filtered temperature field, its long-wavelength ($0 \leqslant k \lesssim 3.5$) and short-wavelength ($k \gtrsim 10$) components prove to be removed.} The resultant pattern (Fig.~\ref{spectrum}b) corresponds to the spectral remainder with wavenumbers in the range $3.5 \lesssim k \lesssim 10$ ({its averaged amplitude is designated as $A_\mathrm f$}; see Figs~\ref{spectrum}c, \ref{spectrum}d) and testifies to the presence of smaller-scale structures. {As can be seen from Fig.~\ref{spectrum}b,} inside the large (primary, or first-scale) structures, there are ones of an intermediate (second) scale and of the smallest (third) scale. The smallest structures appear as light and dark mottles {(which are, however, better distinguishable in Fig.~\ref{moving}b)}. They emerge inside the primary structures and are advected to their periphery by both the large-scale and intermediate-scale flows.

The flow has a continuous spectrum (Fig.~\ref{spectrum}d), which declines with the wavenumber $k$. {While the bright ring in Fig.~\ref{spectrum}a and the main peak in Fig.~\ref{spectrum}d, corresponding to the first-scale structures, are pronounced, the spectral signatures of the second-scale and third-scale structures are smeared and cannot be separated with certainty. Tentatively, the long-wavelength part ($k \lesssim 5$) of the spectrum can be interpreted as a superposition of two peaks -- the already mentioned main one and a minor one located in the range $2.5 \gtrsim k \gtrsim 5$.}

\section{Summary and conclusion}

We have analysed the structure of the velocity and temperature fields found in
our simulations of convection in a plane layer of fluid with
temperature-dependent thermal diffusivity. This dependence is chosen so as to
produce a sharp kink in the static temperature profile near the upper layer
boundary. As a result, the magnitude of the (negative) static temperature
gradient $\rmd T_\mathrm s/\rmd z$ is small over the most part of the layer
thickness but reaches large values in a thin sublayer {near the upper boundary; at the same time, this gradient does not change its sign at any height, and the layer is everywhere convectively unstable}. The random
temperature perturbation introduced in the strongly stratified sublayer
initiates convective motions, which start developing near the upper boundary and then penetrate to progressively deeper layers.

After the initiation of convection, the flow gradually involves the whole layer depth, and the growing size of the
largest structural elements of the velocity field (which we call primary, or first-scale, convection cells) settles down to a relatively steady characteristic value by $t \approx 65\tau_\nu$ (a well-developed, quasi-steady multiscale convection pattern forms by $t \approx 80\tau_\nu$). The primary cells have central upflows and
peripheral downflows.

{Except the first-scale structural elements, ones of two smaller scales can be detected in the flow.} We have identified second-scale structures,
whose existence is evidenced by the presence of ``bridges'', or ``isthmuses'',
intersecting the first-scale cells and the smallest, third-scale structures,
which are advected by the first-scale and second-scale convective flows. While the vertical size of the primary cells corresponds to the whole layer thickness, smaller structures  are localised near the upper boundary. {The cold downdrafts at the borders of the differently scaled cells penetrate to different depths. The smaller the size of the cells, the shallower level their downdrafts pierce. The vertical and the horizontal size are similar for each type of structures, as is typically observed in various known convective flows. The second-scale and third-scale structures are ``suspended'' near the upper layer boundary.}

The simulated flow is a hierarchical superposition of cellular convection structures of three widely different characteristic sizes. This pattern bears
remarkable visual similarities with the pattern of solar convection, which (apart from the existence of giant cells and elusive mini-granules) is a superposition of three cell sorts, viz., supergranules, mesogranules and granules.

It is worth noting that the spatial spectrum of the flow does not directly
indicate the presence of the third-scale structures, {and one also cannot reveal with certainty spectral manifestations of the second-scale structures.} In this
respect, the situation resembles that in the case of solar convection: as we
know, neither mesogranular nor mini-granular scale can be identified in the
power spectra of the velocity field.

{We have thus shown, in the framework of a fairly simple model, that temperature variation of the thermal diffusivity of the fluid can give rise to a scale-splitting effect. This appears to be a nontrivial feature of the hydrodynamics of thermal convection. We did not pursue the aim of reaching a close similarity between the conditions in our model and in the solar convection zone, since the structure of solar convection cannot be accurately reproduced in a layer of an incompressible fluid. However, there are reasons to believe that the detected physical effect of scale splitting due to variations in the thermal diffusivity may have something in common with the structure-forming mechanisms acting in the solar convective flows.} Hopefully, further improvements of the model would make this resemblance more complete. We plan such attempts for the near future.

\section*{Acknowledgements}
We are grateful to V.V. Kolmychkov for developing the computational
code solving the Navier--Stokes equations and for his help in carrying out
the computations. The work of O.V.Shch. and O.S.M. was supported by the Program ``Leading Scientific Schools'' (project no. NSh-9120.2016.2).





\section*{References}

\bibliographystyle{elsarticle-harv}
\bibliography{Shcheritsa}

\begin{thebibliography}{18}
\expandafter\ifx\csname natexlab\endcsname\relax\def\natexlab#1{#1}\fi
\expandafter\ifx\csname url\endcsname\relax
  \def\url#1{\texttt{#1}}\fi
\expandafter\ifx\csname urlprefix\endcsname\relax\def\urlprefix{URL }\fi

\bibitem[{{Abramenko} et~al.(2012){Abramenko}, {Yurchyshyn}, {Goode},
  {Kitiashvili}, and {Kosovichev}}]{Abramenko_etal2012}
{Abramenko}, V.~I., {Yurchyshyn}, V.~B., {Goode}, P.~R., {Kitiashvili}, I.~N.,
  {Kosovichev}, A.~G., Sep. 2012. {Detection of small-scale granular structures
  in the quiet Sun with the New Solar Telescope}. \apjl 756, L27.

\bibitem[{Fletcher(1988)}]{Fletcher}
Fletcher, C. A.~J., 1988. Computational Techniques for Fluid Dynamics. Vol.~2.
  Springer, Berlin.

\bibitem[{{Getling}(1975)}]{Getling:1975}
{Getling}, A.~V., 1975. {Convective motion concentration at the boundaries of a
  horizontal fluid layer with inhomogeneous unstable temperature gradient along
  the height}. Fluid Dyn. 10~(5), 745--750.

\bibitem[{{Getling}(1980)}]{Getling1980}
{Getling}, A.~V., Dec. 1980. {Scales of convective flows in a horizontal layer
  with radiative transfer}. Izv., Atmos. Oceanic Phys. 16, 363--365.

\bibitem[{{Getling}(1998)}]{Getling:1998}
{Getling}, A.~V., 1998. {Rayleigh--B\'enard Convection: Structures and
  Dynamics}. World Scientific; Russian version: URSS, 1999.

\bibitem[{{Getling} et~al.(2013){Getling}, {Mazhorova}, and
  {Shcheritsa}}]{Getling_etal2013}
{Getling}, A.~V., {Mazhorova}, O.~S., {Shcheritsa}, O.~V., Dec. 2013.
  {Concerning the multiscale structure of solar convection}. Geomagn. Aeron.
  53, 904--908.

\bibitem[{Getling and Tikhomolov(2007)}]{GetlingTikh2007}
Getling, A.~V., Tikhomolov, E.~M., 2007. Scale splitting in solar convection.
  In: Trudy XI Pulkovskoi mezhdu\-narodnoi konferentsii po fizike Solntsa
  (Proc. 11th Pulkovo Int. Conf. on Solar Physics). pp. 109--112.

\bibitem[{{Hathaway} et~al.(2000){Hathaway}, {Beck}, {Bogart}, {Bachmann},
  {Khatri}, {Petitto}, {Han}, and {Raymond}}]{Hathaway_etal2000}
{Hathaway}, D.~H., {Beck}, J.~G., {Bogart}, R.~S., {Bachmann}, K.~T., {Khatri},
  G., {Petitto}, J.~M., {Han}, S., {Raymond}, J., Apr. 2000. {The photospheric
  convection spectrum}. \solphys 193, 299--312.

\bibitem[{Kolmychkov et~al.(2006{\natexlab{a}})Kolmychkov, Mazhorova, and
  Popov}]{KMPdu}
Kolmychkov, V.~V., Mazhorova, O.~S., Popov, Y.~P., 2006{\natexlab{a}}. Analysis
  of solution algorithms for the three-dimensional navier-stokes equations in
  the natural variables. Differential Equations 42~(7), 994--1004.

\bibitem[{Kolmychkov et~al.(2006{\natexlab{b}})Kolmychkov, Mazhorova, and
  Popov}]{KMPmm}
Kolmychkov, V.~V., Mazhorova, O.~S., Popov, Y.~P., 2006{\natexlab{b}}. Computer
  simulation for subcritical convection in multi-component alloys. Mathematical
  Modelling and Numerical Analysis 11~(1), 57--71.

\bibitem[{Nordlund et~al.(2009)Nordlund, Stein, and
  Asplund}]{Nrdlnd_Stein_Asplnd_lrsp:2009}
Nordlund, {\AA}., Stein, R.~F., Asplund, M., 2009. Solar surface convection.
  Living Rev. Solar Phys. 6~(2), 2, 1--116.

\bibitem[{{November} et~al.(1981){November}, {Toomre}, {Gebbie}, and
  {Simon}}]{November_etal1981}
{November}, L.~J., {Toomre}, J., {Gebbie}, K.~B., {Simon}, G.~W., May 1981.
  {The detection of mesogranulation on the Sun}. \apjl 245, L123--L126.

\bibitem[{Rieutord and Rincon(2010)}]{RieutRinc2010}
Rieutord, M., Rincon, F., 2010. The sun's supergranulation. Living Rev. Solar
  Phys. 7~(2), 2, 1--82.

\bibitem[{{Sch{\"u}s\-sler}(2013)}]{Schuessler2013}
{Sch{\"u}s\-sler}, M., 2013. {Solar magneto-convection}. In: {Kosovichev},
  A.~G., {de Gouveia Dal Pino}, E., {Yan}, Y. (Eds.), Solar and Astrophysical
  Dynamos and Magnetic Activity, Proc. IAU Symposium No. 294. pp. 95--106.

\bibitem[{Sch\"ussler(2014)}]{Schuessler2014}
Sch\"ussler, M., 2014. private communication\hspace{-2pt}.

\bibitem[{{Shcheritsa} et~al.(2015){Shcheritsa}, {Getling}, and
  {Mazhorova}}]{Shcheritsa_etal2015}
{Shcheritsa}, O.~V., {Getling}, A.~V., {Mazhorova}, O.~S., Feb. 2015.
  {Stratification-induced scale splitting in convection}. Adv. Space Res. 55,
  927--936.

\bibitem[{{Simon} and {Leighton}(1964)}]{SimLeighton1964}
{Simon}, G.~W., {Leighton}, R.~B., Oct. 1964. {Velocity fields in the solar
  atmosphere. III. Large-scale motions, the chromospheric network, and magnetic
  fields.} \apj 140, 1120--1147.

\bibitem[{{V{\"o}gler} et~al.(2005){V{\"o}gler}, {Shelyag}, {Sch{\"u}ssler},
  {Cattaneo}, {Emonet}, and {Linde}}]{Voegler_etal2005}
{V{\"o}gler}, A., {Shelyag}, S., {Sch{\"u}ssler}, M., {Cattaneo}, F., {Emonet},
  T., {Linde}, T., Jan. 2005. {Simulations of magneto-convection in the solar
  photosphere. Equations, methods, and results of the MURaM code}. \aap 429,
  335--351.

\end{thebibliography}



\end{document}